\documentclass[pra,superscriptaddress,groupedaddress]{revtex4}
\usepackage{ae}
\usepackage[T1]{fontenc}
\usepackage[ansinew]{inputenc}
\usepackage{amsmath}
\usepackage{amssymb}
\usepackage[caption=false]{subfig}
\usepackage{multirow}
\usepackage{array}
\usepackage[]{graphicx}
\usepackage{wrapfig}
\usepackage{color}
\usepackage[colorlinks]{hyperref}
\usepackage{lscape}
\begin{document}
\title{Influence of joint measurement bases on sharing network nonlocality}
\author{Amit Kundu}
\email{amit8967@gmail.com}
\affiliation{Department of Applied Mathematics, University of Calcutta, Kolkata- 700009, India}
\author{Debasis Sarkar}
\email{dsarkar1x@gmail.com, dsappmath@caluniv.ac.in}
\affiliation{Department of Applied Mathematics, University of Calcutta, Kolkata- 700009, India}
\begin{abstract}
Sharing network nonlocality in an extended quantum network scenario is the new paradigm in the development of quantum theory. In this paper, we investigate the influence of Elegant joint measurement(in short, EJM) bases in an extended bilocal scenario on sharing network nonlocality via sequential measurement. The work essentially based on the newly introduced[Phys. Rev. Lett. 126, 220401(2021)] bilocal inequality with ternary inputs for end parties and EJM as joint measurement bases in $Alice_n-Bob-Charlie_m$ scenario. Here, we are able to capture all simultaneous violation of this inequality for $(n,m)\in \{(2,1),(1,2),(1,1),(2,2)\}$ cases. We further observe the criteria for sharing network nonlocality where we are able to find also the dependence of the sharing on the amount of entanglement of the joint bases. The effect of the nonlinearity in this inequality is also captured in our results with the symmetrical and asymmetrical violation in this extended scenario. The work will generate further the realization of quantum correlations in network scenario.
\end{abstract}
\date{\today}
\maketitle

\section{Introduction}
Bell Scenario, a trivial picture where two parties are sharing quantum particles from a source and based on their two inputs with two outputs, an inequality called Bell inequality, is being violated, confirms that the correlation they are sharing is nonlocal in nature. The invention of this scenario is a very fundamental and exciting point to start the discussion on quantum nonlocality. In 1964\cite{bellpaper}, John Bell initiated the discussion through the approach where, two parties Alice and Bob share a quantum state from a quantum source with two input(output) choices $A_0$, $A_1$($a \in {\pm 1}$) for Alice and $B_0$, $B_1$($b\in {\pm 1}$) for Bob. Every time Alice and Bob measure one of the inputs randomly and get outcomes $a$ and $b$ randomly and they get a conditional probability distribution $p(a,b|A, B)$ as a correlation. One may assume that a hidden variable $\lambda$ is deciding all the inputs and corresponding outputs for the same scenario and also providing the probability distribution $p_{local}(a,b|A, B,\lambda)$. This correlation $p_{local}(a,b|A, B)$ made by $\lambda$, is a local correlation as it can be made locally without disturbing each other's measurement choices. And ultimately we have an inequality, called Bell inequality, to check the correlation and it turns out that some correlation $p(a,b|A, B)$ from quantum states violate this inequality and become a nonlocal correlation. Now here the measurement choices of Alice and Bob are all sharp or projective measurements(in short, PVM, sometimes called strong measurements) which destroy the nonlocality of a quantum state after one measurement and lead to one of the eigenstates of the measurement basis. In a nonlocality sharing scenario through sequential measurements \cite{sharing1,sharing2,sharing3,sharing4,sharing5,sharing6,sharing7,sharing8,sharing9,sharing10,sharing11,sharing12,sharing13,sharing14,sharing15,sharing16,sharing17,sharing18,sharing19,sharing20}, Alice or Bob or both measure an unsharp measurement(in short, POVM, sometimes called weak measurements) on their respective particles and transfer the particle to the next party sequentially. The unsharp measurement is unsharp in a way that it extracts only that amount of nonlocality which is enough to violate Bell-CHSH inequality and leaves the residual nonlocality in the state and transfer to the next party. In a sequential scenario, the main concern is how many times the particle can be transferred to the next party or how much unsharpness is needed to violate Bell-CHSH inequality between the extended parties.\\

\begin{figure}
\centering
\includegraphics[height=5.8cm,width=9cm]{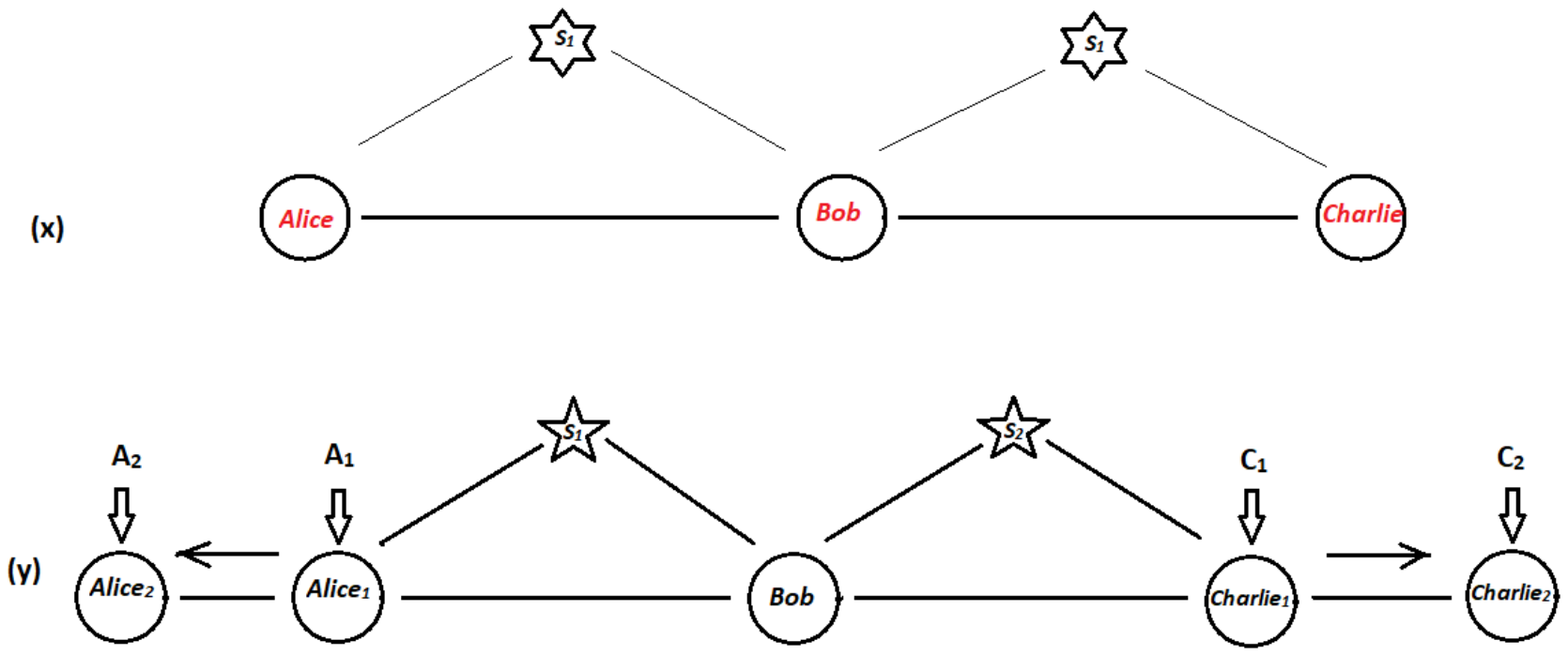}
\caption{(x)Bilocal scenario with two sources. (y)Extended symmetrical bilocal scenario with $Alice_1$ and $Charlie_1$ measuring weak measurement and passing to the $Alice_2$ and $Charlie_2$ respectively.}
\label{fig:BiEx}
\end{figure}

Consider this whole idea in a quantum network scenario, a structure where more than one party connected in a network and more than one quantum source maintaining quantum network nonlocal correlation\cite{review,Kundu20,cava11,cava12,fullNN,GG13,noinbi}. In the trivial quantum network, viz., Bilocal Network \cite{cava11,cava12}, a work has been done recently on sharing standard quantum network nonlocality in an extended Bilocal scenario\cite{sharingBI} with more than one Alice and Charlie. Also, A. Tavakoli et.al. introduced Full Network Nonlocality(in short, FNN) with the help of the general family of partially entangled joint measurement bases called EJM other than Bell States Measurement used by the middle party Bob in the three parties Bilocal scenario\cite{fullNN,tavakoliBI}. They further provided with ternary input choices for Alice and Charlie, a new kind of Bilocal inequality, we will call in this work as TBG inequality. The sitution for using weak measurements are not known to us. In our work, using weak measurement for this ternary systems, we consider an extended Bilocal scenario with two Alices and Two Charlies[Fig:(\ref{fig:BiEx}y)] with measurement EJM for Bob and with the help of the weak measurements of Alices and Charlies except for the two end parties, we try to find whether the nonlocality in the bilocal network can be shared by violating the TBG inequality or not, between $Alice_1-Bob-Charlie_1$, $Alice_1-Bob-Charlie_2$, $Alice_2-Bob-Charlie_1$ and $Alice_2-Bob-Charlie_2$ based on the amount of unsharpness of the sequential measurements. Also we check the dependence of the sharing on the amount of entanglement of the said joint measurement bases incorporated here and we find that only by introducing non-linearity in the TBG inequality the simultaneous violation of those four inequality is possible. The interesting way of dependency of the violation on non-linearity and the noise resistance based on the entanglement of the EJM are also discussed here. These results not only shed light on the characterisation of the network nonlocality but also will provide a path to think about the introduction of the exact amount of the nonlinearity in the TBG inequality. We map this paper following the sections as follows: In section II, we explain the bilocal inequality based on EJM. In section III, we explain how we formalize the process. Section IV and V provides the results. Section VI ended with the Conclusion.

\section{TBG inequality and Full Network Nonlocality}
Consider a scenario with three parties Alice, Bob and Charlie[Fig:(\ref{fig:BiEx}(x))] connected by two independent sources $S_1$ and $S_2$. This is the trivial quantum network structure and the basic scenario of the entanglement swapping where the middle party Bob has two particles and the joint measurement of Bob creates the swapping. Here, the correlation $p(a,b,c|x,z)$ between Alice, Bob and Charlie is said to be a non-bilocal correlation, if the inequality $\sqrt{I} + \sqrt{J} \leq 1$ \cite{cava12}, is violated by the correlation. This inequality is based on the Bell State Measurement(in short, BSM) bases for Bob and the end parties are measuring binary input with binary output. Several work has been done\cite{gisin17,Kundu20,andreoli} on the violation of the bilocal inequality and it is found that only one Bell nonlocal source out of two is enough to violate this bilocal inequality. It is also true for the $n$-local scenario\cite{nlocal}. In\cite{fullNN}, this nonlocality in a quantum network, i.e., network nonlocality is defined as Standard Network Nonlocality(in short, SNN) and the authors introduced Full Network Nonlocality(in short, FNN) where thecorrelation comes only from all Bell nonlocal sources. The SNN can be captured by the inequality $\sqrt{I} + \sqrt{J} \leq 1$, but the FNN cannot be captured by this inequality. Here all the discussions are rounded with the joint measurement bases as BSM. In\cite{25y}, N. Gisin introduced partially entangled measurement bases and showed that these partially entangled bases can reveal nonlocality in complex network as well as in closed loop networks. Later in\cite{tavakoliBI}, Tavakoli et.al. generalized these partially entangled bases with a variable $\theta\in[0, \frac{\pi}{2}]$ which takes the states from EJM to BSM. The generalized bases are,
\begin{equation}\label{equ:EJM}
|\Phi^{\theta}_b\rangle = \frac{\sqrt{3}+e^{i\theta}}{2\sqrt{2}}|\vec{m}_b, -\vec{m}_b\rangle + \frac{\sqrt{3}-e^{i\theta}}{2\sqrt{2}}|-\vec{m}_b, \vec{m}_b\rangle
\end{equation}
where,
\begin{equation}
\begin{split}
&\vec{m_1}=(+1,+1,+1)\\
&\vec{m_2}=(+1,-1,-1)\\ 
&\vec{m_3}=(-1,+1,-1)\\
&\vec{m_4}=(-1,-1,+1)
\end{split}
\end{equation}
are the four vertices of a regular tetrahedron inscribed in a Poincare sphere.
And the states $|-\vec{m}_b\rangle$ are directed to the antipodal direction. Specifically, this tetrahedron vertices can be written in cylindrical coordinates as $\vec{m}_b = \sqrt{3}(\sqrt{1-\eta^{2}_b}cos\phi_b,\sqrt{1-\eta^{2}_b}sin\phi_b,\eta_b)$ and can be defined as
$$|\vec{m}_b\rangle = \sqrt{\frac{1+\eta_b}{2}}e^{-i\phi_b/2}|0\rangle + \sqrt{\frac{1-\eta_b}{2}}e^{i\phi_b/2}|1\rangle$$
Based on these bases they constructed a bilocal inequality(TBG) with the correlator,
$$ S= \sum_{y=z}\langle B^yC_z\rangle - \sum_{x=y}\langle A_xB^y\rangle$$,
$$ T= \sum_{x\neq y\neq z\neq x}\langle A_xB^yC_z\rangle, Z = max(C_{other}),$$
Where $C_{other} $ is the absolute value of all one, two and three party correlators other than those appearing in the expression of $S$ and $T$. The bilocal inequality is,
\begin{equation}\label{equ:BI}
B= \frac{S}{3} - T \leq 3+5Z
\end{equation}
This $Z$ quantity makes the inequality nonlinear. Interestingly the quantum correlation $p_{Q}^{\theta}$ gives the value $Z = 0$. This quantum correlation comes from the two maximally entangled states from two different sources $S_1$ and $S_2$ and the generalized EJM for Bob. The end parties Alice and Charlie here use the ternary measurement choices $x, y, z\in (1,2,3)$ with binary outcomes $a, c\in (+1,-1)$. They also found that the bilocal inequality above, for $Z=0$, is tight in the sense that it constitutes one of the facets of the projection of the $"Z=0"$ slice of the bilocal set correlations onto the $(S,T)$ plane\cite{tavakoliBI}. More accurately, for $Z>0$, they have done the numericals correction term $5Z$ in the bilocal bound $B$[Equ:(\ref{equ:BI})]. This quantum correlation in the bilocal structure based on that generalized EJM has an interesting property called Full Network Nonlocality(in short, FNN). The author showed that this $p_{Q}^{\theta}$ is a FNN correlation except two points at $\theta = 0$ and $\theta = \frac{\pi}{2}$ which also provides the advantages of EJM. So here we can observe a partial entangled measurement bases play an important role in capturing network nonlocality. The effect of this EJM on sharing network nonlocality in an extended Bilocal scenario based on the this TBG inequality needs to be checked. So keeping the ternary inputs choices for end parties in mind and the EJM as joint measurement bases, we investigate, in the next section, how the formalization can be done for sharing network nonlocality in an extended bilocal network with two Alices and two Charlies under weak measurements except for Alice and Charlie at end by simultaneous violation of the inequality for $(n,m)\in \{(2,1),(1,2),(1,1),(2,2)\}$ [Equ:(\ref{equ:BI})].
\section{Formalization of extended bilocal scenario}
As illustrated in the figure[fig:\ref{fig:BiEx}(y)], here we will discuss an extended bilocal network. In a three parties bilocal scenario, Alice, Bob and Charlie are connected by two sources $S_1$ and $S_2$ in a way that $S_1$ generates two-qubit quantum states and distributes it through Alice and Bob and similarly $S_2$ distributes through Bob and Charlie. In a symmetric way, we are adding one Alice and one Charlie to both ends and this time each observer $Alice_n$($Charlie_m$) chooses ternary observables, denoted by $X_n \in (A_{n,i})$ for Alices and $Z_m \in (C_{m,j})$ for Charlies with binary outcomes $a_n \in (a_{n,k})$ and $c_{m}\in (c_{m,l})$, where $n, m, k, l \in {1,2}$, and $i, j\in{1,2,3}$ and $a_{n,i}, c_{m,j} \in \{0,1\}$. Here Bob performs Elegant Joint Measurement bases from(\ref{equ:EJM}) defined as $Y$ with four distinguishable outcomes $b$ with the corresponding vector $m_b$. The joint probability distribution of the outcomes of the observables can be described as $P(a_1, a_2, b, c_1, c_2|X_1, X_2, Y, Z_1, Z_2)$.
\\

From this extended scenario we can get the joint probability distribution of any structure $Alice_n-Bob-Charlie_m$, defined as,
\begin{equation}\label{equ:prob}
P(a_n, b, c_m|X_n, Y, Z_m) = \frac{1}{4}
\sum_{a_{n'},c_{m'},X_{n'},X_{m'}}P(a_1, a_2, b, c_1, c_2|X_1, X_2, Y, Z_1, Z_2)
\end{equation}
In this scenario $Alice_1$ and $Charlie_1$ will perform weak measurements, whereas Bob, $Alice_2$ and $Charlie_2$ will carry out strong measurements.
Now, we will consider the probability distribution $P(a_n, b, c_m|X_n, Y, Z_m)$ for any $Alice_n-Bob-Charlie_m$ in the extended bilocal scenario based on the weak, strong and joint measurements. If the two sources $S_1$ and $S_2$ are emitting two states $\rho_{AB}$ and $\rho_{BC}$ respectively, the whole state of the system can be derived by,
\begin{equation}
\rho_{ABC} = \rho_{AB}\otimes\rho_{BC}
\end{equation}
Bob performs Elegant Joint Measurements on the two particles he receives with four possible outputs $b$. The observers, $Alice_1$ and $Alice_2$, will measure their shared particle sequentially, similarly for $Charlie_1$ and $Charlie_2$ on the other side. The first observer of each side performs the optimal weak measurements, while the second observer of each side will do strong measurements. The observables of each observer can be defined as,
\begin{equation}
\begin{split}
A_{1,1} = C_{1,1} = \cos\beta_1\cos\gamma_1\sigma_x + (\sin\alpha_1\sin\beta_1\cos\gamma_1 - \cos\alpha_1\sin\gamma_1)\sigma_y + (\cos\alpha_1\sin\beta_1\cos\gamma_1 + \sin\alpha_1\sin\gamma_1)\sigma_z\\
A_{1,2} = C_{1,2} = \cos\beta_1\sin\gamma_1\sigma_x + (\sin\alpha_1\sin\beta_1\sin\gamma_1 + \cos\alpha_1\cos\gamma_1)\sigma_y + (\cos\alpha_1\sin\beta_1\sin\gamma_1 - \sin\alpha_1\cos\gamma_1)\sigma_z\\
A_{1,3} = C_{1,3} = -\sin\beta_1\sigma_x + \sin\alpha_1\cos\beta_1\sigma_y +\cos\alpha_1\cos\beta_1\sigma_z.
\end{split}
\end{equation}
Similarly for $Alice_2$ and $Charlie_2$, the angles will be $\alpha_2$, $\beta_2$ and $\gamma_2$. $\sigma_x$, $\sigma_y$ and $\sigma_z$ are the Pauli matrices. \\

When Bob performs the Elegant Joint measurement on the two particle with the result $b_0b_1\in \{00,01,10,11\}$, the state will be,
\begin{equation}
\rho_{ABC}^{b_0b_1} = (\mathbb{I}\otimes\rho_{b_0b_1}\otimes\mathbb{I}).\rho_{ABC}.(\mathbb{I}\otimes\rho_{b_0b_1}\otimes\mathbb{I})^{\dagger}.
\end{equation}
The reduced state on Alice's and Charlie's side can be obtained by tracing over the Bob's part,
\begin{equation}
\rho_{AC}^{b_0b_1} = tr_{B}(\rho_{ABC}^{b_0b_1})
\end{equation}
Here, $\rho_{AC}^{b_0b_1}$ is not normalized. After the weak measurement of $Alice_1$, the reduced state will be,
\begin{equation}
\rho_{X_1}^{a_1} = \frac{F_1}{2}\rho_{AC}^{b_0b_1} + \frac{1 + (-1)^{a_1}G_1 - F_1}{2}[(U^1_{X_1}.\rho_{AC}^{b_0b_1}.(U^1_{X_1})^{\dagger}] + \frac{1 - (-1)^{a_1}G_1 - F_1}{2}[(U^0_{X_1}.\rho_{AC}^{b_0b_1}.(U^0_{X_1})^{\dagger}]
\end{equation}
where $U^{a_n}_{X_n} = \prod^{a_n}_{X_n}\otimes\mathbb{I}$ and $\prod^{a_n}_{X_n}\in {\frac{\mathbb{I}+(-1)^{a_{n,i}}A_{n,i}}{2}}$.
$F_1$ is the quality factor which represents the undisturbed extent to the state of $Alice_1$'s qubit after she measured and $G_1$ is the precision factor which quantifies the information gain from $Alice_1$'s measurements. $Alice_2$ performs a strong measurement $X_2$ with the outcome $a_2$, the state will be,
\begin{equation}
\rho^{a_2}_{X_2} = U^{a_2}_{X_2}\rho^{a_1}_{X_1}(U^{a_2}_{X_2})^{\dagger}.
\end{equation}
Similarly, $Charlie_1$ performs weak measurements $Z_1$ on his qubit with the quality factor $F_2$ and precision factor $G_2$ of the measurements. The reduced state, when the outcome is $c_1$ can be described as 
\begin{equation}
\rho_{Z_1}^{c_1} = \frac{F_2}{2}\rho_{X_2}^{a_2} + \frac{1 + (-1)^{c_1}G_2 - F_2}{2}[(U^1_{Z_1}.\rho_{X_2}^{a_2}.(U^1_{Z_1})^{\dagger}] + \frac{1 - (-1)^{c_1}G_2 - F_2}{2}[(U^0_{Z_1}.\rho_{X_2}^{a_2}.(U^0_{Z_1})^{\dagger}],
\end{equation}
where $U^{c_m}_{Z_m} = \mathbb{I}\otimes\prod^{c_m}_{Z_m}$ and $\prod^{c_m}_{Z_m}\in {\frac{\mathbb{I}+(-1)^{c_{m,j}}C_{m,j}}{2}}$. And at last $Charlie_2$ performs a strong measurement $Z_2$ with the outcome $c_2$, the reduced state will be,
\begin{equation}
\rho^{c_2}_{Z_2} = U^{c_2}_{Z_2}\rho^{c_1}_{Z_1}(U^{c_2}_{Z_2})^{\dagger}.
\end{equation}
So from the above analysis, we get the joint probability distribution $P(a_1, a_2, b, c_1, c_2|X_1, X_2, Y, Z_1, Z_2) = Tr[\rho_{Z_2}^{c_2}]$. In this whole process the measurements of all observers are completely independent and unbiased. From the eq:\ref{equ:prob}, we can get the probability distribution for any $Alice_2-Bob-Charlie_m$ or $Alice_n-Bob-Charlie_2$. Here simply we can say that the sequence of $Bob - Alice_1 - Alice_2 - Charlie_1 - Charlie_2$ is maintained through the measurement process.
\\ 

To check the correlation between any combination of $Alice_n-Bob-Charlie_m$ we need to check the $B_{nm}$ from equ:\ref{equ:BI} and its simultaneous violations confirm the existence of the network nonlocal correlation, we are now going to search here that at what conditions the simultaneous violations can happen. 

\section{Network nonlocal sharing in extended bilocal scenario based on ejm}
\begin{figure}
     \centering
     \begin{minipage}[c]{0.45\linewidth}
         \centering
         \includegraphics[width=\textwidth]{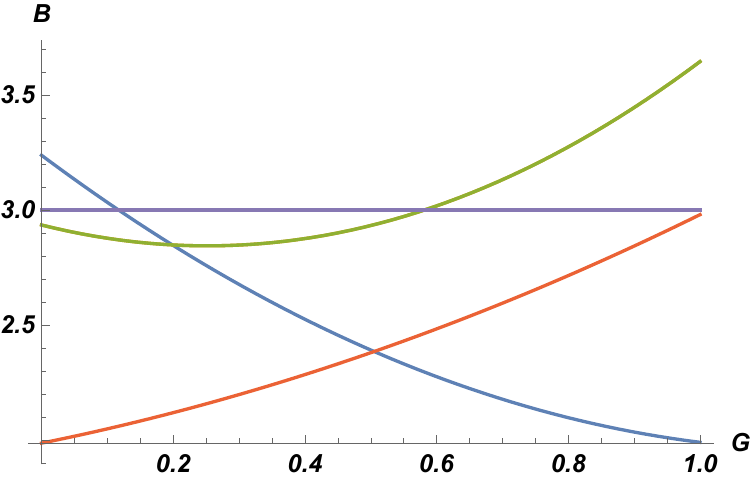}
         \label{fig:F+G}
     \end{minipage}
     \hfill
     \begin{minipage}[c]{0.45\linewidth}
         \centering
         \includegraphics[width=\textwidth]{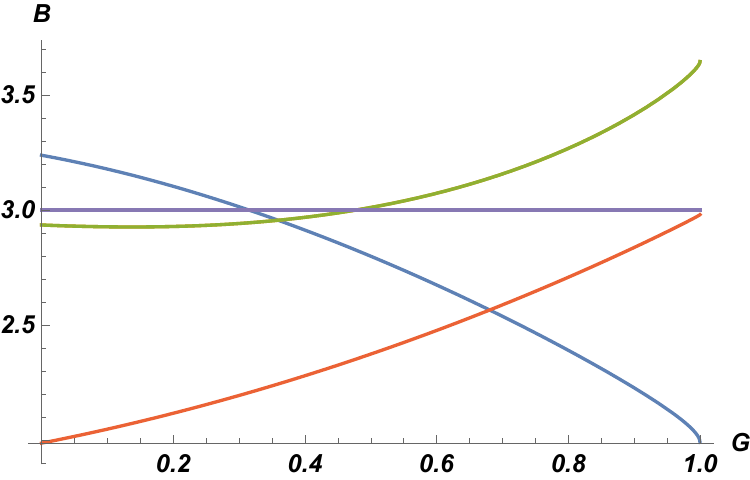}
         \label{fig:FF+GG}
     \end{minipage}
        \caption{The left one corresponds the TBG value for $\theta = 0$ with $F+G = 1$, and the right one corresponds the TBG value for $\theta = 0$ with $F^2+G^2 = 1$. The green line represents $B_{12}$ and $B_{21}$. The red line represents $B_{11}$ and the blue line is $B_{22}$.}
        \label{fig:Z0T0}
\end{figure}
\begin{figure}
     \centering
     \begin{minipage}[c]{0.45\linewidth}
         \centering
         \includegraphics[width=\textwidth]{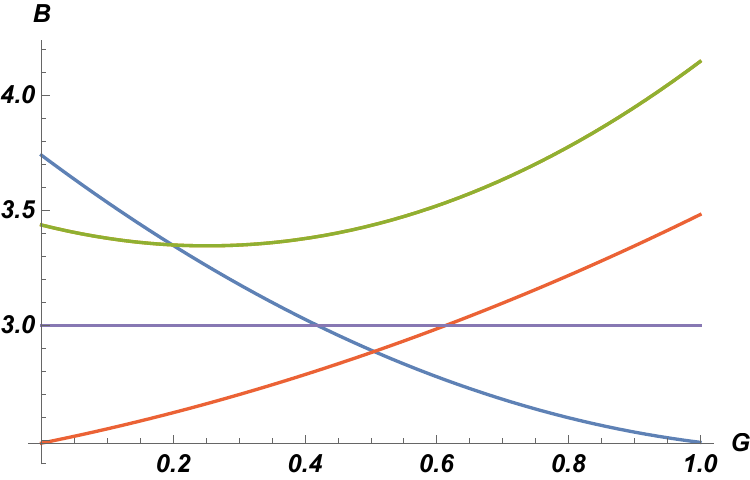}
         \label{fig:Z0.5}
     \end{minipage}
     \hfill
     \begin{minipage}[c]{0.45\linewidth}
         \centering
         \includegraphics[width=\textwidth]{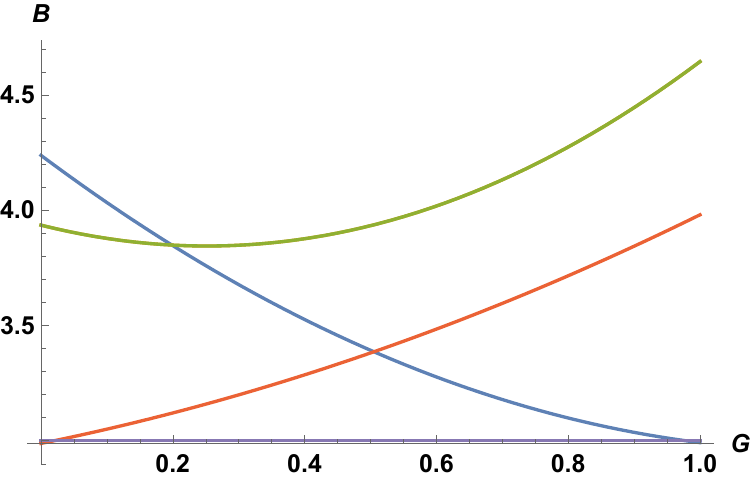}
         \label{fig:Z0.6}
     \end{minipage}
        \caption{Varying $Z$, we get the TBG violation like that where $\theta = 0$ with $F+G = 1$ for $Z = 0.5$ and $Z = 0.6$ respectively.}
        \label{fig:ZN}
\end{figure}
\begin{figure}
     \centering
     \begin{minipage}[c]{0.45\linewidth}
         \centering
         \includegraphics[width=\textwidth]{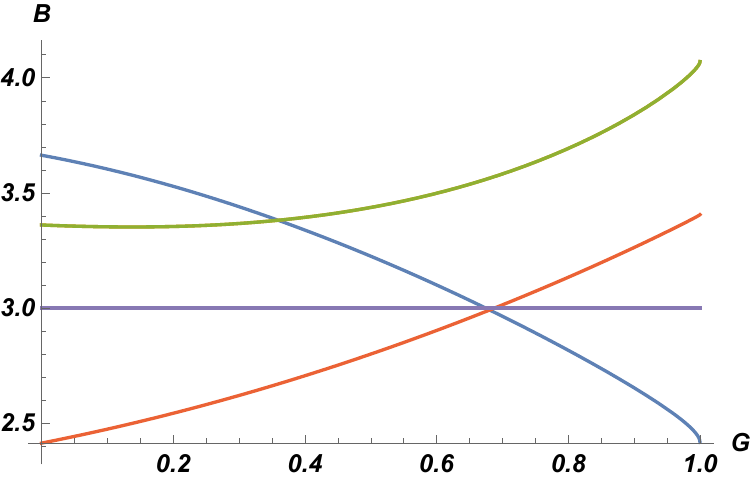}
         \label{fig:Z0.485}
     \end{minipage}
     \hfill
     \begin{minipage}[c]{0.45\linewidth}
         \centering
         \includegraphics[width=\textwidth]{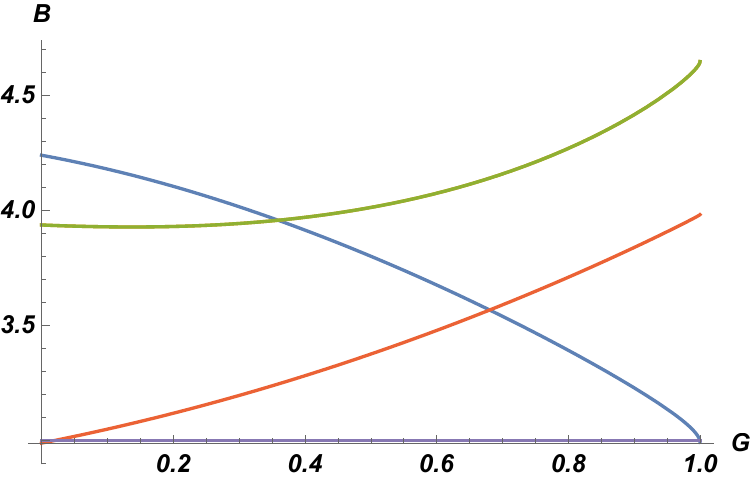}
         \label{fig:ZZ0.6}
     \end{minipage}
        \caption{Respectively $Z = 0.485$ and $Z = 0.6$, we get the TBG violation like that where $\theta = 0$ with $F^2+G^2 = 1$.}
        \label{fig:ZNS}
\end{figure}

Now assume that the sources $S_1$ and $S_2$ send pair of particles in the maximally entangled state, $\rho_{AB} = \rho_{BC} = |\psi\rangle\langle\psi|$, $|\psi\rangle = \frac{1}{\sqrt{2}}(|01\rangle - |10\rangle)$. Next, Alice and Chalie's particles end up in the corresponding states depending upon the Bob's result measuring EJM(Equ:\ref{equ:EJM}). Here our aim is to check whether the network nonlocality sharing happens between the two sides in the scenario(Fig:\ref{fig:BiEx}). Here all the measurement choices of $Alice_1$ and $Charlie_1$ are totally random and we only mean to achieve the maximum violation of the TBG inequality. $Alice_1$ and $Charlie_1$ will choose the optimal measurement to violate TBG inequality between $B_{11}$ and similarly for other. The optimal settings we get $\gamma_1 = \gamma_2 = 0$, $\alpha_1 = \alpha_2 = \beta_1 = \beta_2 = \frac{\pi}{4}$ for all Alices and Chalrlies. For the weak measurements, the pointer distributions are two, optimal and square pointer distribution, with the quality factor $F^2 + G^2 = 1$ and $F + G = 1$ respectively with $G \in[0,1]$. We want a simultaneous violation of all the TBG quantity $B_{11}$, $B_{12}$, $B_{21}$ and $B_{22}$ based on the $F$ and $G$ with chosen $F_1 = F_2 = F$ and $G_1 = G_2 = G$ while middle party Bob will do single EJM measurement, whose entanglement depends on the variable $\theta$. Surprisingly, the four quantity does not violate simultaneously anywhere in the range of $G\in[0,1]$ and $\theta\in[0,\frac{\pi}{2}]$ for both pointer and square distribution while $Z = 0$. In\cite{tavakoliBI}, they made the TBG inequality more tighter, introducing more nonlinearity, for $Z>0$ and we also have some interesting results for $Z>0$(fig:\ref{fig:ZN}) with $F + G = 1$ and with $F^2 + G^2 = 1$(fig:\ref{fig:ZNS}). For square pointer distribution the simultaneous violation of TBG inequality start from $Z = 0.525$ and for optimal distribution the value will be $Z = 0.485$.

As we use the general Elegant Joint Measurement for $Bob$, we can vary the $\theta$ from the measurement bases and we get that for $\theta = \frac{\pi}{2}$ no network nonlocality sharing is happening. In table(\ref{table:theta}), the conditions on $Z$ based on $\theta$ are denoted.
\begin{center}
\begin{tabular}{|c|c|c|c|c|c|c|c|c|c|c|}
\hline
For & $\theta$ & $0$ & $\frac{\pi}{8}$ & $\frac{\pi}{4}$ & $\frac{3\pi}{8}$ & For & $0$ & $\frac{\pi}{8}$ & $\frac{\pi}{4}$ & $\frac{3\pi}{8}$\\
\hline
$F + G = 1$ & $Z$ & 0.525 & 0.532 & 0.555 & 0.579 & $F^2 + G^2 = 1$ & 0.485 & 0.5 & 0.535 & 0.573\\
\hline
\end{tabular}
\label{table:theta}
\end{center}
With this results we can make the statement that a certain amount of non-linearity can make the TBG inequality tighter which provides the novel forms of network nonlocal correlation. With the changing theta the change of this non-linearity is also captured here.
\\

In this quantum network scenario the correlation is always effected by the noisy state production from the quantum sources. In this network sharing process in extended bilocal scenario, discussion on noise resistance is also a very interesting aspect. If the two sources share Werner states($\rho_W$) instead of maximally entangled states with noise parameter $v_i$ of the form,
\begin{equation}\label{w}
\rho_w = v_i(|\psi\rangle\langle\psi|) + \frac{1-v_i}{4}\mathbb{I}
\end{equation}
$v_1$ and $v_2$ are the noise parameters for $\rho_{AB}$ and $\rho_{BC}$.
Maintaining all the previous assumptions intact we have the conditions to have simultaneous violation of the TBG inequality and these are,
\begin{center}
\begin{tabular}{|c|c|c|c|}
\hline
$\theta$ & $0$ & $\frac{\pi}{8}$ & $\frac{\pi}{4}$\\
\hline
$Z$ & 0.58 & 0.578 & 0.575\\
\hline
$V$ & 0.45 & 0.71 & 0.82\\
\hline
\end{tabular}
\end{center} 
Here the value of $V = \sqrt{v_1v_2}$ is the critical visibility for sharing and the value confirms at which point the simultaneous violation is possible for $Alice_1-Bob-Charlie_1$ and $Alice_2-Bob-Charlie_2$. 
\section{Discussion and conclusion}
Two major things discussed here are the introduction of weak measurements in a symmetrical extended Bilocal scenario in a sequential form for the ternary inputs systems and the effect of the partially entangled joint measurement bases with the changing entanglement for the middle party, Bob. We are able to show here how the joint measurement bases are important in a quantum network scenario not just to reveal the type of correlations but also use it in an experimental information processing tasks in a complex quantum network scenario. Using EJM bases the nonlocality sharing in an extended Bilocal scenario is a step towards a more practical way to understand the network correlations. Here we are not just checking the amount of entanglement in joint measurement bases to share network nonlocality with weak measurements but also providing a point to think of modification of the TBG inequality with proper non-linearity. We here discussed the range in which the simultaneous violation of the TBG inequality can happen for all four $B_{11}$, $B_{12}$, $B_{21}$ and $B_{22}$ cases which is extremely helpful for experimental realisation. Also, the impact of the pointer types on the sharing is brought here. The noise resistance is also discussed based on the angle $\theta$ and the non-linearity $Z$. Lastly, we observe that this is an area of interest for further research and the results are very important in the sense to characterise the exact form of the correlations distributing in an extended quantum network with two different, independent sources, with ternary weak measurements and EJM and also in observing experimentally,  sharing network nonlocality based on EJM. In\cite{tavakoliBI}, they proposed an experimental realization of EJM which can be implemented to observe this sharing. Also making ternary POVM is possible in the laboratory. The whole range of results can be useful for the network nonlocality theoretically and practically.

\section{Acknowledgements}
 The authors AK and DS acknowledge the work as part of QUest initiatives by DST India.

\end{document}